\documentclass[pra,aps,showpacs,twocolumn]{revtex4}
\usepackage{graphics,bm}
\usepackage{amssymb,ulem}
\usepackage{epsfig}
\usepackage{epsf}
\usepackage[usenames]{color}
\usepackage{latexsym}
\usepackage{amsfonts}
\usepackage{xcolor,MnSymbol}

\begin{document}

\title{$2k_F$-Friedel to $4k_F$-Wigner oscillations in the one-dimensional Fermi gases under confinement}
\author{Gao Xianlong}
\email{gaoxl@zjnu.edu.cn}
\affiliation{Department of Physics, Zhejiang Normal University, Jinhua, Zhejiang Province, 321004, China}

\date{\today}

\begin{abstract}

Density oscillations of confined one-dimensional Fermi gases of contact repulsive interactions in a continuous space are discussed within Bethe-ansatz based spin-density functional theory. The results are compared against the exact analytical and the exact diagonalization method. For unpolarized system, the number of the peaks in the density profiles is doubled signaling the crossover of the $2k_F$-Friedel to $4k_F$-Wigner oscillations (with $k_{\rm F}$ the Fermi wave vector). For both unpolarized and polarized systems, a threshold of the short-range interaction strength can be found where $N_f$-peak Wigner oscillations of $4k_F$ wave vector appear in the density profile ($N_f$ is the total particle number).

\end{abstract}
\pacs{03.75.Ss,71.15.Mb,71.10.Pm}
\maketitle

\section{Introduction}
In higher dimensions a true crystal, named Wigner crystal, where the density is perfectly ordered, is possible when the repulsion between particles dominates over the kinetic energy in the case of small density. In one dimension, quantum fluctuations prevent any type of true spin-symmetry breaking and long-range order in the ground state, and thus the crystal does not have perfect order~\cite{ref:giamarchi_book,Schulz,Wigner}. However, in one dimension, the equilibrium positions of the electrons are
assumed to be equally spaced and a solid like state can be produced when the correlation length exceeds the sample length~\cite{ref:Mueller}. The liquid-to-quasi-Wigner crystal crossover when the density decreasing can be characterized by the onset of the $4k_F$-peak in the static structure factor following from the Luttinger liquid predictions~\cite{Schulz, ref:Shulenburger}, with $k_{\rm F}$ the Fermi wave vector (related to the fermion density $n$ by $k_{\rm F}=\pi n/2$), or by the $N_f-1$ high-contrast peaks in the pair correlation function~\cite{ref:Wang}, with $N_f$ the particle number in the system.

In the system where the constituent particles are not electrons but dilute atoms, the fundamental interaction is no longer
of long-range Coulomb type but of short-range contact one, which is the case in the field of ultracold atomic systems.
In this system, the ground-state density profile displays at strong coupling well-pronounced Wigner oscillations with a `$4k_{\rm F}$ periodicity'. However, the static structure factor or the pair correlation function does not show any signature of Wigner-molecule-type correlations like that in the system of long-range interaction~\cite{Astrakharchik0},
which means Wigner-type molecules are absent in the system of short-range interaction~\cite{ref:Wang}. In this paper, we focus only on the ``$2k_F\rightarrow 4k_F$ crossover" in the density profiles: for vanishing and weak interactions the ground state is liquid-like with $2k_{\rm F}$-Friedel oscillations, whereas at strong repulsions a periodicity of $4 k_{\rm F}$-Wigner oscillation emerges.

For the one-dimensional (1D) harmonically trapped system of spin-balanced or imbalanced fermionic atomic gases, many numerical schemes have been used in studying lots of interesting quantum effects~\cite{Machida,Molina,Rigol,Astrakhardik,Casula,exact diagonalization,Nikkarila}.
A density-functional scheme using as reference the Bethe-ansatz solvable homogeneous system was proposed to calculate the ground-state properties~\cite{ref:burke,ref:gao_pra_2006}. However, this approach was found to fail at the strong-coupling limit~\cite{ref:gao_pra_2006}. More precisely, the exchange-correlation potential proposed there is not able to describe the $2k_{\rm F}\rightarrow 4k_{\rm F}$ crossover, which, as noted above, is expected to occur upon increasing the strength of the repulsive interactions between antiparallel-spin particles.

A simple functional is proposed in Ref.~[\onlinecite{ref:saeed_pra_2007}] for a two-particle system that embodies this crossover and is capable of describing inhomogeneous Luttinger liquids with strong repulsions. The main idea is to capture the tendency to $4k_F$-Wigner oscillations by adding an infinitesimal spin-symmetry-breaking magnetic field to the Hamiltonian and by resorting to spin-density functional theory (SDFT) based on the spin-polarized Bethe-ansatz equations~\cite{ref:Capelle,ref:gaoAsgari}.

For fermions loaded in a 1D optical lattice, the system can be modeled by a 1D Fermi-Hubbard model.
Friedel oscillations of wavelength $2k_{\rm F}$ and $4k_{\rm F}$ component are analyzed in this model with open boundary condition by both the Bethe-ansatz based SDFT, the exact diagonalization~\cite{ref:Vieira}, bosonization, and numerical density matrix renormalization group calculations~\cite{1D_wigner_HM,soffing_pra_2011,ref:Bedurftig}. It was found that the $2k_{\rm F}\rightarrow 4k_{\rm F}$ crossover takes place at $U=4t$. For an unpolarized system, if a tiny magnetization which is proportional to the difference of the spin-up and spin-down densities is induced by hand into the exchange-correlation potential, the $4k_F$-oscillation appears in the density profile at the expense of violating the Lieb-Mattis theorem, while for a polarized system, SDFT can reproduce both the $2k_{\rm F}$ and $4k_{\rm F}$ oscillations without violating the Lieb-Mattis theorem~\cite{ref:Vieira}. It is found that the $4k_F$ density oscillations is strongly related to the large Fermi sea induced by the strong repulsion between different spin species~\cite{1D_wigner_HM,soffing_pra_2011}.

Some 1D systems, for example, the 1D Gaudin-Yang gas in a homogeneous continuum space, or the 1D Lieb-Wu model of particles in a discrete lattice, i.e., 1D Fermi-Hubbard model, where the fermions interact through attractive or repulsive short-range interactions, are exactly solvable. However, in the inhomogeneous case only some special systems can be solved exactly. For example, the 1D two-component Fermi gas in an infinite potential well can be solved exactly by the Bethe-ansatz techniques, where the ground state properties are investigated~\cite{ref:Wei}.
The density profiles are evaluated from weak to very strong interactions. In the strongly interacting limit, the density profile tends to that of the spinless fermions though the momentum distributions differ. An exact analytical solution of 1D spin-1/2 fermions with infinite repulsion for arbitrary confining potential is obtained by the combination of Girardeau's hard-core contacting boundary condition and group theoretical method~\cite{ref:Guan}. For the spin-imbalanced case they found that the spin-up and spin-down density distributions show alternative peaks avoiding overlapping together and the parity symmetry.

In this paper, we will employ the Bethe-ansatz based SDFT to investigate the 1D inhomogeneous system with general interaction strengths. Our results are compared against the above mentioned exact analytic and the exact numerical diagonalization method. The Wigner oscillations while increasing coupling strength for both unpolarized and polarized cases will be studied.

The rest of the paper is organized as follows. In Sec. II, we present the model and the method. The numerical results are followed in Sec. III. In Sec. IV, we conclude what we found.

\section{Model and the method}
In this work we consider a 1D system of two-component Fermi gas with a contact repulsive interaction of strength $g_{\rm 1D}$~\cite{ref:olshanii_1998}.
The two fermion species are taken to have the same mass $m$, with $N_\uparrow$ spin-up species and $N_\downarrow$ spin-down ones. The total number of particles is $N_f=N_\uparrow+N_\downarrow$. The system is described by the Hamiltonian,
\begin{equation}\label{eq:igy}
{\cal H}=-\frac{\hbar^2}{2m}\sum_{i=1}^{N_f}\frac{\partial^2}{\partial z^2_i}+g_{\rm 1D}\sum_{i=1}^{N_{\uparrow}}\sum_{j=1}^{N_{\downarrow}}\delta(z_i-z_j)+V_{\rm ext}\,,
\end{equation}
where contact interactions between parallel-spin particles are suppressed by the Pauli exclusion principle and $V_{\rm ext}$ the confining potential. We consider that the atoms are either confined in a 1D box of length $L$ ($V_{\rm ext}=0$ inside the box and $V_{\rm ext}=+\infty$ elsewhere) or subjected to a strongly anisotropic harmonic potential $V_{\rm ext}=\sum_{i=1}^{N} V_{\rm ext}(z_i)=(m\omega^2_{\|}/2)\sum_{i=1}^{N}z^2_i$, characterized by angular frequencies $\omega_\perp$ and $\omega_\|$ in the radial and axial directions with $\omega_\| \ll \omega_\perp$. For the system of harmonic trap,
we choose the harmonic-oscillator length $a_{\|}=\sqrt{\hbar/(m\omega_{\|})}$ as unit of length and the harmonic-oscillator quantum $\hbar\omega_{\|}$ as unit of energy.
The Hamiltonian (\ref{eq:igy}) can be shown to be governed by the dimensionless coupling parameter
\begin{equation}\label{eq:lambda}
\lambda=\frac{g_{\rm 1D}}{a_{\|}\hbar\omega_{\|}}\,.
\end{equation}
For $V_{\rm ext}=0$ the Hamiltonian (\ref{eq:igy}) reduces to the homogeneous Gaudin-Yang model solvable by means of the Bethe-ansatz technique~\cite{ref:GY} and determined by
the linear density $n=N/L$, by the spin polarization $\zeta=(N_{\uparrow}-N_{\downarrow})/N$, and by the interaction strength $g_{\rm 1D}$.
\begin{figure}
\begin{center}
\includegraphics[width=1.00\linewidth]{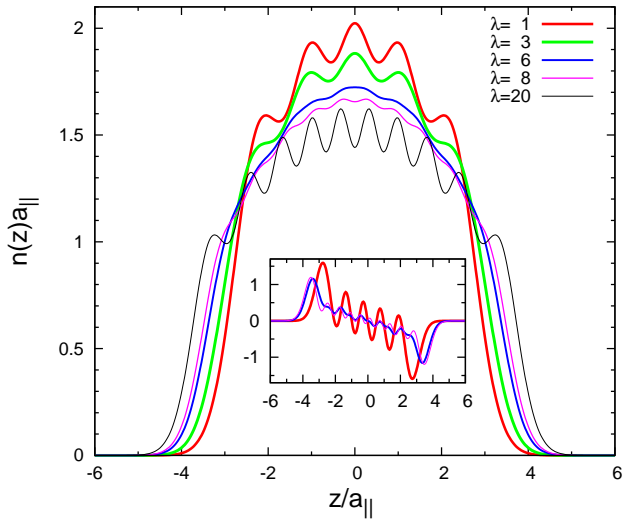}
\caption{(Color online) The ground-state density distributions as a function of $z/a_{||}$ for a balanced Fermi gas with $N_\uparrow=N_\downarrow=5$
confined in a harmonic trap for various interaction strengths. The crossover from $2k_F$- to $4k_F$-oscillations with increasing
interaction strength $\lambda$ is showed by the doubling of the crests in the density profile.
The dominance of the $4k_F$-oscillations over $2k_F$-ones can be better evidenced by the density derivative $dn(z)/dz$.
In the inset, the derivative of the density with respect to $z$ is shown for $\lambda=1, 6$, and $8$ as an indicator to examine the $2k_{\rm F}\rightarrow 4k_{\rm F}$ crossover.
\label{fig:one}}
\end{center}
\end{figure}
\begin{figure}
\begin{center}
\includegraphics[width=1.00\linewidth]{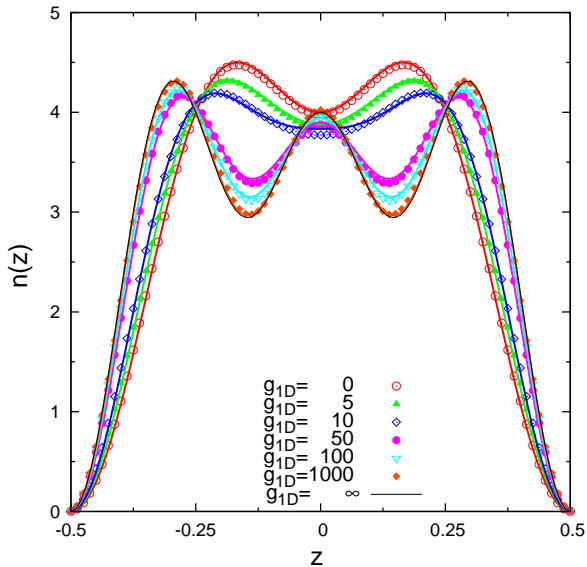}
\caption{(Color online) The ground-state density distributions (in arbitrary length$^{-1}$ unit) as a function of $z$ (in arbitrary length unit) for a polarized Fermi gas with $N_\uparrow=2$ and $N_\downarrow=1$
confined in a box of length $L=1$ for various interaction strengths. The SDFT results are plotted in symbols and the exact Bethe-ansatz ones in (colored) lines. The thin (black) solid line represents the result in the infinite repulsive interaction limit.
\label{fig:three}}
\end{center}
\end{figure}

The ground-state spin-density distributions $n_\sigma(z)$ can be calculated within SDFT by solving self-consistently the Kohn-Sham (KS) equation,
\begin{equation}\label{eq:kss}
\left[-\frac{\hbar^2}{2m}\frac{\partial^2}{\partial z^2}+V^{(\sigma)}_{\rm KS}[n_\sigma](z)\right]\varphi_{\alpha,\sigma}(z)=\varepsilon_{\alpha,\sigma}\varphi_{\alpha,\sigma}(z)
\end{equation}
together with the closure
\begin{equation}\label{eq:closure}
n_\sigma(z)=\sum_{\alpha=1}^{N_\sigma}\left|\varphi_{\alpha,\sigma}(z)\right|^2\,.
\end{equation}
Here, $V^{(\sigma)}_{\rm KS}[n_\sigma](z)=V^{(\sigma)}_{\rm H}[n_\sigma](z)+V^{(\sigma)}_{\rm xc}[n_\sigma](z)+V_{\rm ext}(z)$ is the spin-dependent effective KS potential.
The first term in $V^{(\sigma)}_{\rm KS}$ is the mean-field term $V^{(\sigma)}_{\rm H}=g_{\rm 1D}n_{\bar \sigma}(z)$, while the second term is the exchange-correlation potential defined as the functional derivative of the exchange-correlation energy $E_{\rm xc}[n_\sigma]$ evaluated at the ground-state density profile, $V^{(\sigma)}_{\rm xc}=\delta E_{\rm xc}[n_\sigma]/\delta n_\sigma(z)|_{\rm \scriptscriptstyle GS}$.

To obtain the density distribution $n_\sigma(z)$ from Eqs.~(\ref{eq:kss}) and~(\ref{eq:closure}), approximations to the exchange-correlation functions $E_{\rm xc}$ are unavoidable. The local-spin-density approximation (LSDA) is known to provide an excellent description of the ground-state properties of a variety of inhomogeneous systems~\cite{ref:Giuliani_and_Vignale}. In the following we employ a Bethe-ansatz-based LSDA (BALSDA) functional for the exchange-correlation potential,
\begin{eqnarray}\label{eq:balda}
E_{\rm xc}[n_\sigma] & \rightarrow & E^{\rm LSDA}_{\rm xc}[n_\sigma]\\
&=&\int dz\, n(z)\left.\varepsilon^{\rm hom}_{\rm xc}(n,\zeta,g_{\rm 1D})\right|_{n\rightarrow n(z),\zeta\rightarrow \zeta(z)}\,,\nonumber
\end{eqnarray}
where the exchange-correlation energy $\varepsilon^{\rm hom}_{\rm xc}$ of the homogeneous Gaudin-Yang model is defined by
\begin{eqnarray}\label{eq:xc}
\varepsilon^{\rm hom}_{\rm xc}(n,\zeta,g_{\rm 1D})&=&\varepsilon_{\rm GS}(n,\zeta,g_{\rm 1D})-\kappa(n,\zeta)\nonumber\\
&-&\varepsilon_{\rm H}(n,\zeta,g_{\rm 1D})\,.
\end{eqnarray}
Here $\varepsilon_{\rm GS}(n,\zeta,g_{\rm 1D})$ is the ground-state energy of the exact Bethe-ansatz solution of the model,
$
\varepsilon_{\rm H}(n,\zeta,g_{\rm 1D})=\frac{1}{4}g_{\rm 1D}n^2(1-\zeta^2)
$
is the mean-field energy, and
$\kappa(n,\zeta)=\pi^2\hbar^2 n^2(1+3\zeta^2)/24m$ is the noninteracting kinetic energy~\cite{ref:GY}.
\begin{figure*}
\begin{center}
\tabcolsep=0 cm
\begin{tabular}{cc}
\scalebox{0.35}[0.35]{\includegraphics{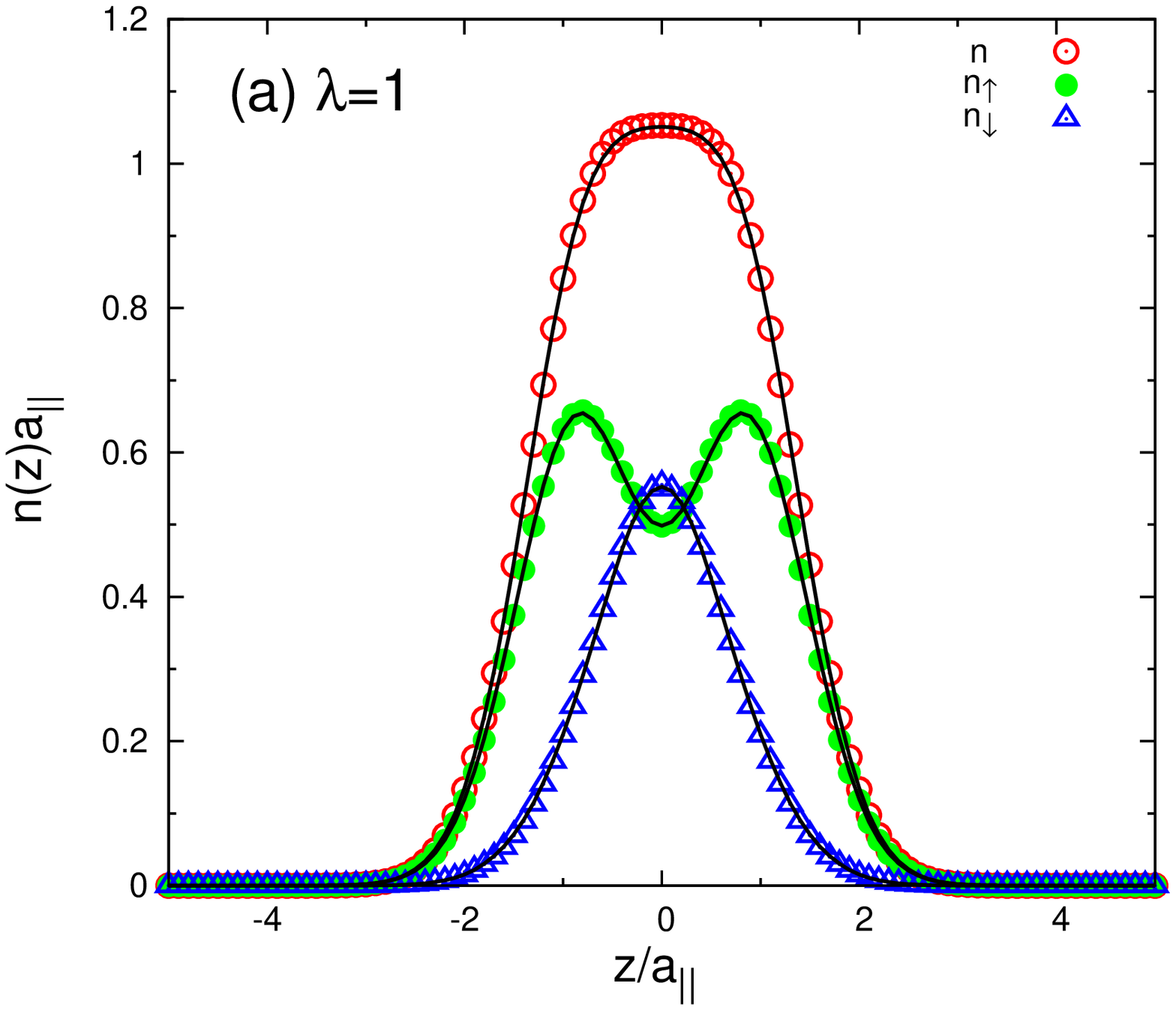}}&
\scalebox{0.35}[0.35]{\includegraphics{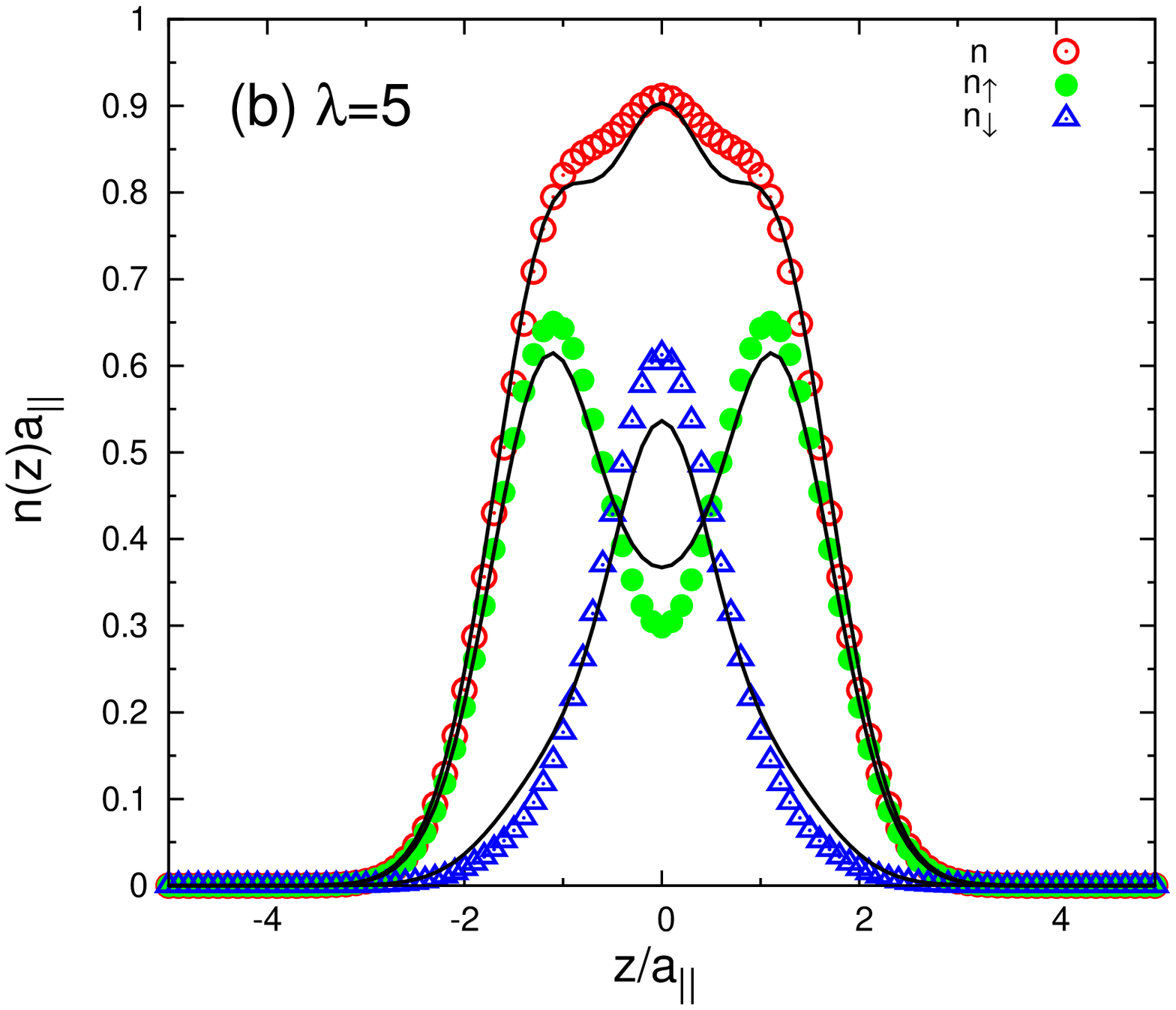}}\\
\scalebox{0.35}[0.35]{\includegraphics{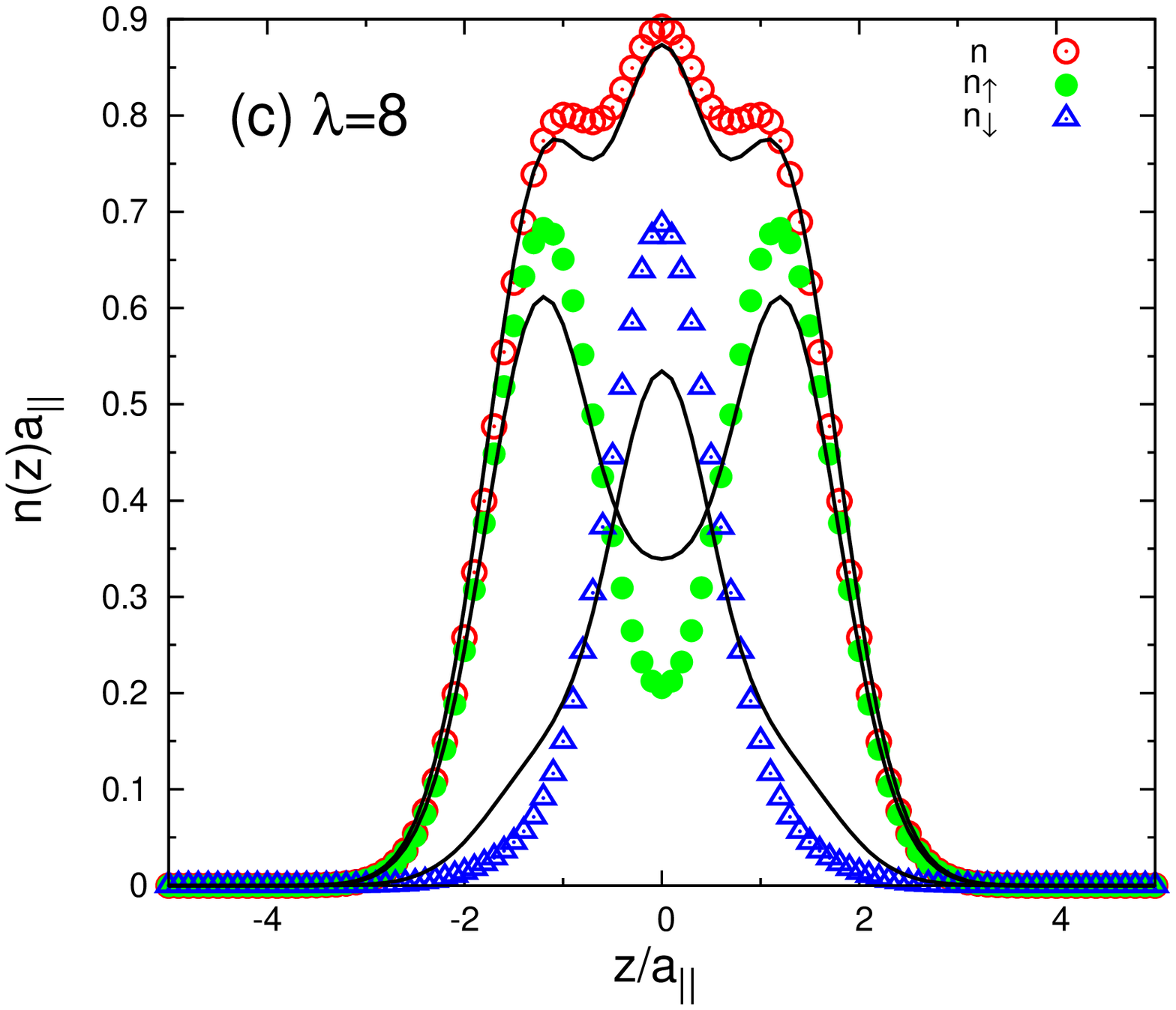}}&
\scalebox{0.35}[0.35]{\includegraphics{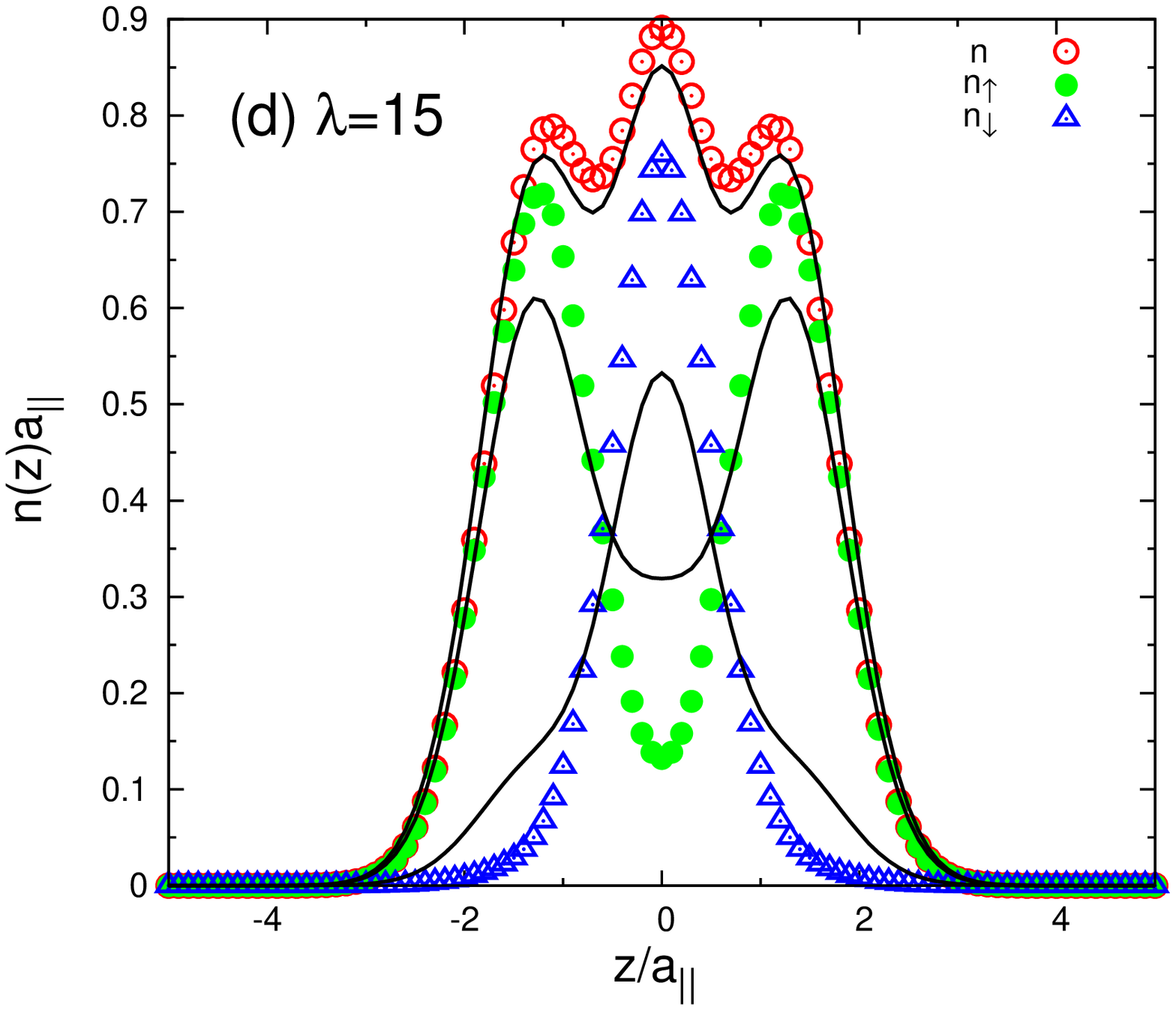}}
\end{tabular}
\caption{(Color online) The ground-state density distributions as a function of $z/a_{||}$ for a polarized Fermi gas with $N_\uparrow=2$ and $N_\downarrow=1$
confined in a harmonic trap ranges from weak to strong interaction strengths: (a) $\lambda=1$; (b) $\lambda=5$; (c) $\lambda=8$; (d) $\lambda=15$. The SDFT results are plotted in symbols ($\color{red}\circ$, $ \color{green}\bullet$, and $\color{blue}\triangle$, respectively, for $n$, $n_\uparrow$, and $n_\downarrow$ ) and the exact diagonalization ones in lines.
\label{fig:four}}
\end{center}
\end{figure*}
\begin{figure}
\begin{center}
\tabcolsep=0 cm
\includegraphics[width=1.00\linewidth]{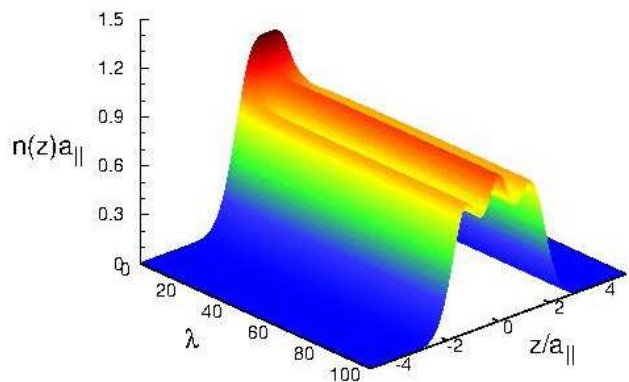}
\caption{(Color online) The 3D plot of the ground-state density distributions as a function of the dimensionless interaction strength $\lambda$ and $z/a_{||}$ for a polarized Fermi gas with $N_\uparrow=2$ and $N_\downarrow=1$
confined in a harmonic trap. The results are obtained based on the SDFT.
\label{fig:five}}
\end{center}
\end{figure}

\section{Numerical results}
In this section, we present our numerical results by employing the Bethe-ansatz based SDFT to investigate the 1D inhomogeneous system with broad interaction strengths. Our results are compared against the exact numerical diagonalization method while this method is available when the system is small. For the polarized system in a box, we compare our results with the analytical ones which are available in this case by the exact Bethe-ansatz method.

\subsection{Unpolarized system confined in a trap}
In Fig. \ref{fig:one}, we study an unpolarized Fermi gas with the number of total particles $N_f=10$
confined in a harmonic trap. The ground-state density distributions is calculated for various interaction strengths within SDFT.
For the unpolarized system, the KS equations have to be solved self-consistently by taking one of the following methods to obtain
the correct $2k_{\rm F}\rightarrow 4k_{\rm F}$ crossover. (i) One can either add an infinitesimal spin-symmetry-breaking magnetic field.
The field strength is infinitesimal small and reduced to zero during the self-consistent solution of the KS equations. (ii)
Alternatively one can solve the KS equations from an initial guess which has slightly broken spin symmetry [{\it i.e.} an initial guess with  $n_\uparrow(z)\neq n_\downarrow(z)$], without using any $B$ field, (iii) The last possibility is to break the symmetries of the density distributions just one time by shifting the center of the densities during the self-consistent cycle [for example, let $n_\uparrow(z)= n_\uparrow(z+b)$ with $b$ a nonzero value].
When $\lambda$ is small, the above mentioned methods restore the results from the Bethe-ansatz based local density approximation (BALDA)~\cite{ref:Capelle}.
However when $\lambda$ is large, the final result for the total density $n(z)$ from the BALSDA is very different from that resulting from the BALDA theory~\cite{ref:gao_pra_2006}. We want to mention here that, without using the above mentioned symmetry-breaking procedure, the BALSDA theory produces the same results as that of BALDA. That is, for the BALSDA theory, to achieve a correct shape for the density profile, one has to use one of the (i)-(iii) procedure~\cite{ref:saeed_pra_2007}. This method has long been used in the literature in discussing the ground-state of the $H_2$ molecule at any bond length. The symmetry-breaking choice typically produce a spin-unpolarized density out to a critical bond length by localizing an up-spin electron on one nucleus and a down-spin electron on the other. While this leads to an incorrect spin-density, but it leads to the correct dissociation limit for the energy and the correct total density~\cite{Gunnarsson0}.

At the weak interaction regime, the density is characterized by $2k_F$-Friedel oscillations with $N_f/2$ peaks.
As the interaction strength increases to the strongly interacting limit, the number of peaks in the density profile is doubled. The atoms are distributed in order to minimize the interparticle repulsion energy. This leads to the emergence of $N_f$ crests in the density profile characterized by strong correlations and the $4k_F$-Wigner oscillations. In this case the atoms behave like a fully polarized (or noninteracting spinless) fermions as in the Tonks-Girardeau gas~\cite{ref:hard-core boson}.
Numerically we find that a crossover from the $2k_F$-Friedel oscillations of $N_f/2$ peaks to the $4k_F$-Wigner oscillations happens around $\lambda=6.2$ accompanied of $N_f$ peaks. The crossover is smooth, see, for example, the extensive density-matrix renormalization group study by S\"{o}ffing {\it et al.}~\cite{1D_wigner_HM,soffing_pra_2011}. In the inset, a derivative of the density with respect to $z/a_{\|}$ is used as an indicator to signal the $2k_{\rm F}\rightarrow 4k_{\rm F}$ crossover. We find more clearly that $N_f$ peaks appear after the threshold value.

\begin{figure}
\begin{center}
\includegraphics[width=1.00\linewidth]{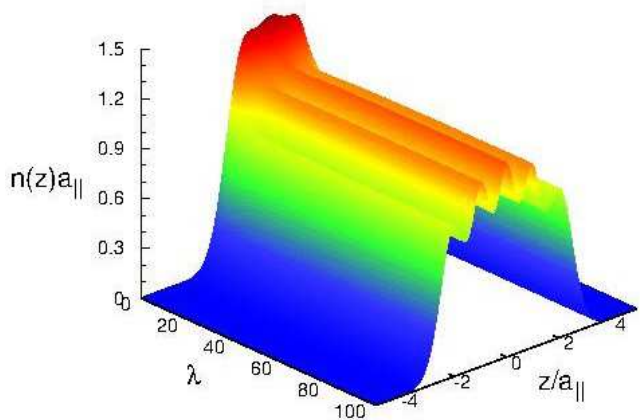}
\caption{(Color online) The 3D plot of the ground-state density distributions as a function of the dimensionless interaction strength
$\lambda$ and $z/a_{||}$ for a polarized Fermi gas with $N_\uparrow=3$ and $N_\downarrow=2$
confined in a harmonic trap. The results are obtained based on the SDFT.
\label{fig:six}}
\end{center}
\end{figure}
\begin{figure}
\begin{center}
\includegraphics[width=1.00\linewidth]{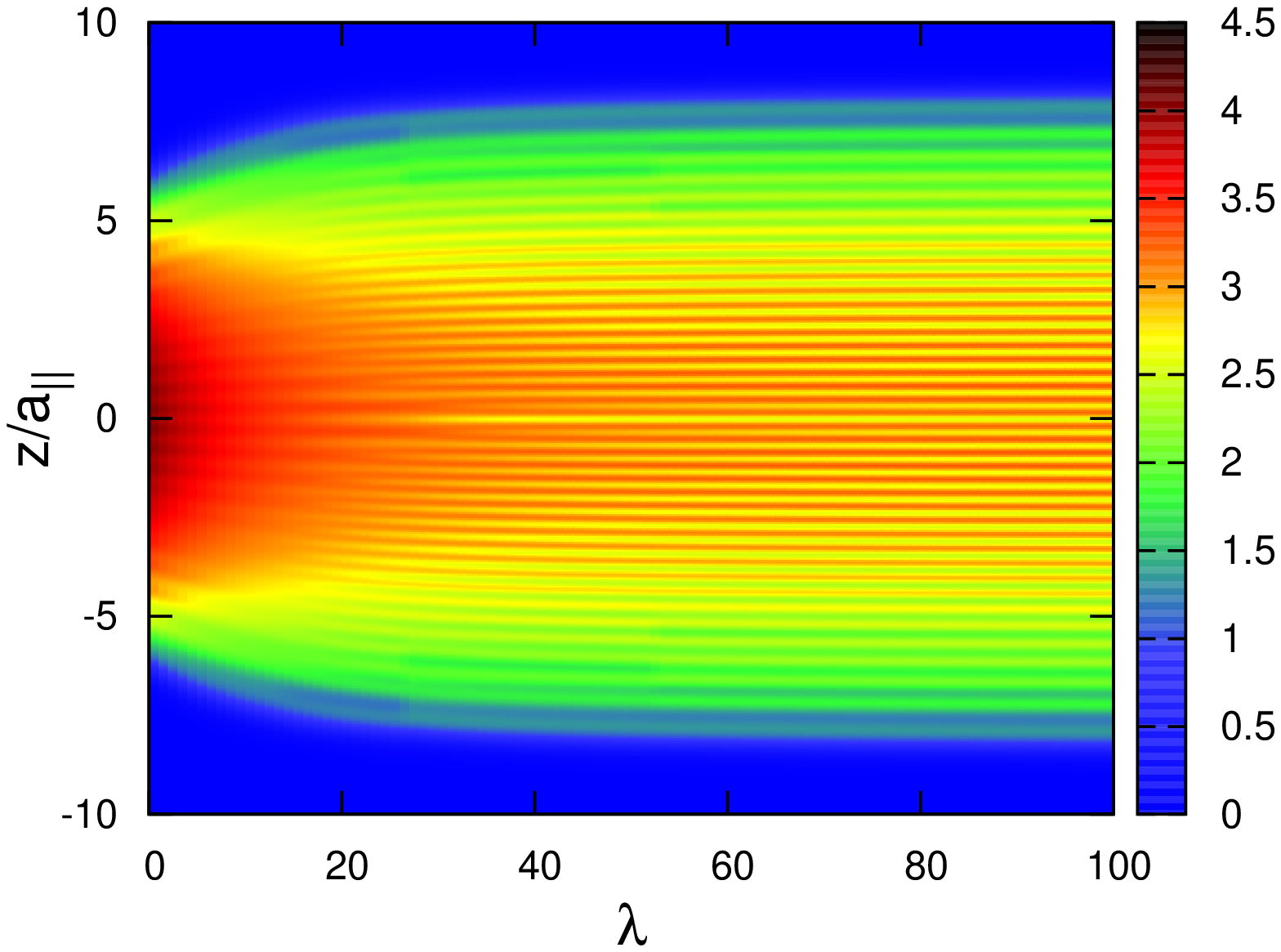}
\caption{(Color online) The 2D contour plot of the ground-state density distributions as a function of the dimensionless interaction strength
$\lambda$ and $z/a_{||}$ for a polarized Fermi gas with $N_\uparrow=22$ and $N_\downarrow=18$
confined in a harmonic trap. The results are obtained based on the SDFT.
\label{fig:eight}}
\end{center}
\end{figure}
\begin{figure}
\begin{center}
\includegraphics[width=1.00\linewidth]{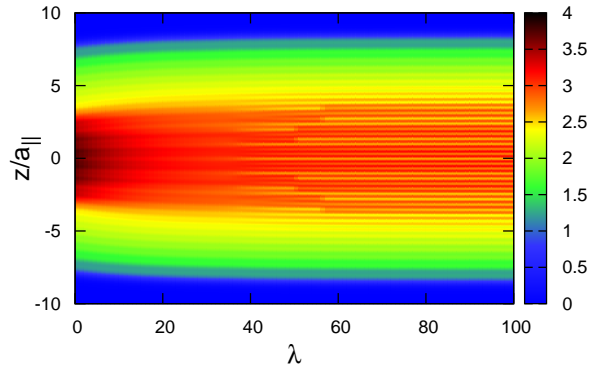}
\caption{(Color online) The 2D contour plot of the ground-state density distributions as a function of the dimensionless interaction strength
$\lambda$ and $z/a_{||}$ for a polarized Fermi gas with $N_\uparrow=34$ and $N_\downarrow=6$
confined in a harmonic trap. The results are obtained based on the SDFT.
\label{fig:nine}}
\end{center}
\end{figure}
In order to evaluate the credibility of the present method, we compare our results in the following against exact methods
for a polarized system in a box (in \ref{sec1}) and a harmonically confined system
of small number of Fermi atoms (in \ref{sec2}).

\subsection{Polarized system in a box}\label{sec1}

In this section we consider a polarized Fermi gas with $N_\uparrow$ up-spin fermions and $N_\downarrow$ down-spin fermions confined in a box of length $L$, which is exactly solvable by the Bethe-ansatz method~\cite{ref:Wei}. In Fig. \ref{fig:three}, we illustrate the ground-state density distributions as a function of $z$ based on the Bethe-ansatz based SDFT for $N_\uparrow=2$ and $N_\downarrow=1$. $L$ is chosen as units of length.
The interaction strength sets the relative importance of the kinetic energy and the interaction. For weak interaction strength, $2$ peaks appear in the density.
When interaction strength becomes stronger (around $g_{1D}=15$), $3$ peaks emerge. At even stronger interaction strength, exchange is suppressed and the atoms act as though they were noninteracting and spinless in analogy to the case studied for bosons with an infinite contact repulsion~\cite{ref:hard-core boson}. For $g_{1D}=1000$, the infinite repulsive interaction limit is reached, which is compared to the spinless fermions results.
In general, when the interaction strength is large enough, the density $n(z)$ distributes between $N_f$ maxima, which is different from the confined 1D electrons of long-range Coulomb interactions where the density vanishes almost completely between well separated maxima~\cite{ref:Jauregwi,1D_wigner}.
We find, in almost all the range of the interaction strength, SDFT gives good results comparing against the exact Bethe-ansatz ones.
The good agreement may be that the exact exchange-correlation energy of the homogeneous system has almost the same form as that of the inhomogeneous (hard-wall) one which is also integrable.

\subsection{Polarized system confined in a trap}\label{sec2}

Here we report our studies on the ground-state density distributions for a polarized Fermi gas confined in a harmonic trap. The SDFT results are compared against the ones by the exact diagonalization method~\cite{ref:Guan}. In Fig. \ref{fig:four} we plot the density profiles for the system with $N_\uparrow=2$ and $N_\downarrow=1$ for $\lambda=1, 5, 8$, and $15$. We find that for $\lambda < 2$, both the total ground-state densities $n(z)$ and the spin-resolved densities show very good agreement between the SDFT results and the exact diagonalization ones. When $\lambda\gtrsim 2$, SDFT still keeps the overall shape of the total ground-state densities well. However, the respective amplitudes of the spin-up and spin-down density are exaggerated by the SDFT. Even for this weakness, the SDFT can help us qualitatively understand the $2k_{\rm F}\rightarrow 4k_{\rm F}$ crossover.
For weak interaction strength, the interaction term gives smaller contribution comparing to the kinetic term, such that $n(z)$ is
dominated by the lowest occupied single-particle states. For the system we studied here, one flat peak is found (See Fig. \ref{fig:four}(a) and (b)).
For strong interaction strength $\lambda\gtrsim 3.8$~\cite{ref:notes}, a structure consisting of $N_f=3$ peaks emerges (See Fig. \ref{fig:four}(c) and (d)), that is, $n(z)$ is signaled in finite region by $N_f=3$ maxima. For even stronger interaction strength $\lambda\gtrsim 15$, the total atom distribution can be described by spinless fermions and the spins tend to decouple with exponentially small exchange interactions. The system degenerates with ferromagnetic and anti-ferromagnetic states~\cite{ref:Guan}.

In Figs. \ref{fig:five} and \ref{fig:six}, the 3D plots of the ground-state density distributions are shown as a function of the dimensionless interaction strength $\lambda$ and $z/a_{||}$. The transition from $2k_{\rm F}$-Friedel oscillations to $4 k_{\rm F}$-Wigner oscillations can be judged by the change in the peaks in the total density distributions, which vary from $1$ to $3$ in Fig. \ref{fig:five} happened at $\lambda\gtrsim 3.8$, and from $3$ to $5$ in Fig. \ref{fig:six} happened at $\lambda\gtrsim 4.7$.

At strongly repulsive limit, the polarized system shows the same properties as those of the unpolarized one, where the atoms become impenetrable and behave as they were noninteracting and spinless.

DFT provides a method through which we can access larger systems beyond the reach of QMC and DMRG simulations.
In the following we study the polarized system confined in a trap with particle number of $N_f=40$,
which is not accessible by means of the present exact diagonalization method. Contour plots are shown in Figs. \ref{fig:eight} and \ref{fig:nine}, respectively, for $p=0.1$ and $0.7$. A clear signature of $2k_{\rm F}\rightarrow 4k_{\rm F}$ crossover is signaled by the emergence of $N_f$ peaks at the
strong interaction strength.

From the above discussion, we find that, in spite of the short-range nature of the interaction, density profile shows pronounced well-separated $N_f$ peaks at strong repulsive interactions.
However, the densities between the peaks do not drop to zero, which is different from the 1D system of long-range Coulomb interaction, where they drop almost to zero in some sense resembling a charge density wave and a quasi-Wigner molecule.

\section{Conclusions}
In this work, we study a 1D system of two-component Fermi gases with a contact repulsive interaction.
The unpolarized or polarized atoms are either confined in a 1D box or subjected to a strongly anisotropic harmonic potential.
The total and spin-resolved densities are calculated within the Bethe-ansatz based spin-density functional theory.
The results are compared to those from the exact Bethe-ansatz method and the exact diagonalization scheme. We found that the Bethe-ansatz based SDFT gives a faithful description for the ground state properties at weak interaction strength but exaggerates the amplitude of the density distributions especially for the corresponding spin-up and spin-down densities at strong interaction strength. In spite of the short-range nature of the interaction, a pronounced $N_f$-peak structure is observed in the density profile in the strongly interacting regime.

For unpolarized systems, the density is characterized by $2k_F$-Friedel oscillations with $N_f/2$ peaks at the weak interaction regime.
As the interaction strength increases, the number of peaks in the density profile is doubled.
This leads to the emergence of $4k_F$-Wigner oscillations of $N_f$ peaks in the density profile. Numerically we found a threshold interaction strength $\lambda_c$ at which the $2k_F\rightarrow 4k_F$ transformation happens. For polarized systems, at weak interaction strength, the peaks in the density profile depend on the system parameters. However, when increasing the interaction strength,
a threshold value can be found where $N_f$-peak oscillations appear in the density profile signaling the $4k_F$-Wigner oscillations.
At strongly repulsive limit, the total energy and the total density profile of the system own the same properties, as those of
the fully unpolarized one. There, the atoms become impenetrable and behave as they were noninteracting and spinless.

{\it Acknowledgments}
This work was supported by NSF of China under grant nos. 11174253 and 10974181, Zhejiang Provincial Natural Science Foundation
under grant no R6110175. It is a pleasure to thank Giovanni Vignale, Marco Polini, Shu Chen, and Wei Li for several useful discussions.


\begin{thebibliography}{99}
\bibitem{ref:giamarchi_book}
	    T. Giamarchi, {\it Quantum Physics in One Dimension} (Clarendon Press, Oxford, 2004).
\bibitem{Schulz}
        H. J. Schulz, Phys. Rev. Lett. {\bf 64}, 2831 (1990); {\bf 71}, 1864 (1993).
\bibitem{Wigner}
        E. Wigner, Phys. Rev. {\bf 46}, 1002 (1934).
\bibitem{ref:Mueller}
        Erich J. Mueller, Phys. Rev. B {\bf 72}, 075322 (2005);
        G. A. Fiete, J. Qian, Y. Tserkovnyak, and B. I. Halperin, Phys. Rev. B {\bf 72}, 045315 (2005).
\bibitem{ref:Shulenburger}
        L. Shulenburger, M. Casula, G. Senatore, and R. M. Martin, Phys. Rev. B {\bf 78}, 165303 (2008).
\bibitem{ref:Wang}
        J.-J. Wang, W. Li, S. Chen, G. Xianlong, M. Rontani, and M. Polini, arXiv:1202.3284, accepted to publish in Phys. Rev. B
\bibitem{Astrakharchik0}
        G.E. Astrakharchik and Yu. E. Lozovik, Phys. Rev. A {\bf 77}, 013404 (2008);
        G.E. Astrakharchik and M.D. Girardeau, Phys. Rev. B {\bf 83}, 153303 (2011).
\bibitem{Machida}
        M. Machida, S. Yamada, Y. Ohashi, and H. Matsumoto, Phys. Rev. A {\bf 74}, 053621 (2006);
        M. Machida, S. Yamada, M. Okumura, Y. Ohashi, and H. Matsumoto, Phys. Rev. A {\bf 77}, 053614 (2008).
\bibitem{Molina}
        R. A. Molina, J. Dukelsky, and P. Schmitteckert, Phys. Rev. Lett. {\bf 99}, 080404 (2007).
\bibitem{Rigol}
        M. Rigol, A. Muramatsu, G. G. Batrouni, and R. T. Scalettar, Phys. Rev. Lett. {\bf 91}, 130403 (2003);
        M. Rigol and A. Muramatsu, Phys. Rev. A {\bf 69}, 053612 (2004);
        Opt. Commun. {\bf 243}, 33 (2004).
\bibitem{Astrakhardik}
        G. E. Astrakharchik, D. Blume, S. Giorgini, and L. P. Pitaevskii, Phys. Rev. Lett. {\bf 93}, 050402 (2004).
\bibitem{Casula}
        M. Casula, D. M. Ceperley, and E. J. Mueller, Phys. Rev. A {\bf 78}, 033607 (2008);
        N. Helbig, J.I. Fuks, M. Casula, M.J. Verstraete, M.A.L. Marques, I.V. Tokatly, and A. Rubio, Phys. Rev. A {\bf 83}, 032503 (2011).
\bibitem{exact diagonalization}
        T. Husslein, W. Fettes, and I. Morgenstern, Int. J. Mod. Phys. C {\bf 8}, 397 (1997).
\bibitem{Nikkarila}
        J.-P. Nikkarila, M. Koskinen, S. M. Reimann, and M. Manninen, New J. Phys. {\bf 10}, 063013 (2008);
        J.-P. Nikkarila, M. Koskinen, and M. Manninen, Eur. Phys. J. B {\bf 64}, 95 (2008).
\bibitem{ref:burke}
        R.J. Magyar and K. Burke, Phys. Rev. A {\bf 70}, 032508 (2004).
\bibitem{ref:gao_pra_2006}
	   G. Xianlong, M. Polini, R. Asgari, and M.P. Tosi, \pra {\bf 73}, 033609 (2006).
\bibitem{ref:saeed_pra_2007}
        S.H. Abedinpour, M. Polini, G. Xianlong, and M.P. Tosi, Phys. Rev. A {\bf 75}, 015602 (2007).
\bibitem{ref:Capelle}
        N.A.~Lima, M.F.~Silva, L.N.~Oliveira, and K.~Capelle,
    	Phys. Rev. Lett. {\bf 90}, 146402 (2003);
    	K.~Capelle, N.A.~Lima, M.F.~Silva, and L.N.~Oliveira, in The
    	Fundamentals of Electron Density, Density Matrix and Density
    	Functional Theory in Atoms, Molecules and Solids, Kluwer series,
    	``Progress in Theoretical Chemistry and Physics,'' edited by
    	N.I.~Gidopoulos and S.~Wilson (Kluwer, Dordrecht, 2003).
\bibitem{ref:gaoAsgari}
        G. Xianlong and R. Asgari, \pra {\bf 77}, 033604 (2008).
\bibitem{ref:Vieira}
        D. Vieira, H.J.P. Freire, V.L. Campo Jr., and K. Capelle, J. Mag. Mag. Mat. {\bf 320}, E418 (2008).
\bibitem{1D_wigner_HM}
        S. A. S\"{o}ffing, M. Bortz, I. Schneider, A. Struck, M. Fleischhauer, and S. Eggert, Phys. Rev. B {\bf 79}, 195114 (2009).
\bibitem{soffing_pra_2011}
	    S. A. S\"{o}ffing, M. Bortz, and S. Eggert, \pra {\bf 84,} 021602(R) (2011).
\bibitem{ref:Bedurftig}
        G. Bed\"urftig, B. Brendel, H. Frahm, and R. M. Noack, Phys. Rev. B {\bf 58}, 10225 (1998).
\bibitem{ref:Wei}
        B.-B. Wei, J.-P. Cao, S.-J. Gu, H.-Q. Lin, arXiv:0807.2154v1.
\bibitem{ref:Guan}
        L. Guan, S. Chen, Y. Wang, and Z.-Q. Ma, Phys. Rev. Lett. {\bf 102}, 160402 (2009).
\bibitem{ref:olshanii_1998}
	   M. Olshanii, \prl {\bf 81}, 938 (1998);
	   T. Bergeman, M.G. Moore, and M. Olshanii, {\it ibid.} {\bf 91}, 163201 (2003).	
\bibitem{ref:GY}
	   M. Gaudin, Phys. Lett. {\bf 24A}, 55 (1967); C.N. Yang, \prl {\bf 19}, 1312 (1967).
\bibitem{ref:Giuliani_and_Vignale}
	   G.F. Giuliani and G. Vignale, {\it Quantum Theory of the Electron Liquid} (Cambridge University Press, Cambridge, 2005).
\bibitem{Gunnarsson0}
       O. Gunnarsson and B.I. Lundqvist, Phys. Rev. B {\bf 13}, 4274 (1976);
       J.I. Fuks, A. Rubio, and N.T. Maitra, Phys. Rev. A {\bf 83}, 042501 (2011).
\bibitem{ref:hard-core boson}
    L. Tonks, Phys. Rev. {\bf 50}, 955 (1936);
    M. Girardeau, J. Math. Phys. {\bf 1}, 516 (1960);
    B.~Paredes, A.~Widera, V.~Murg, O.~Mandel, S.~F\"olling,
    I.~Cirac, G.V.~Shlyapnikov, T.~W. H\"ansch, and I.~Bloch, Nature
    {\bf 429}, 277 (2004); T.~Kinoshita, T.~Wenger, and D.S.~Weiss,
    Science {\bf 305}, 1125 (2004).
\bibitem{ref:Jauregwi}
    K. Jauregwi, W. H\"{a}usler, and B. Kramer, Europhys. Lett. {\bf 24} 581 (1993).
\bibitem{1D_wigner}
    B. Szafran, F.M. Peeters, S. Bednarek, T. Chwiej, and J. Adamowski, Phys. Rev. B {\bf 70}, 035401 (2004);
    A. Secchi and M. Rontani, Phys. Rev. B {\bf 80}, 041404 (2009).
\bibitem{ref:notes}
    For the polarized system confined in a harmonic trap with $N_{\uparrow}=2$ and $N_{\downarrow}=1$, the SDFT based on BALSDA gives the threshold
    interaction strength at $\lambda_c=3.8$ where the crossover from $2k_F$-Friedel to $4k_F$-Wigner oscillations happens, while the exact diagonalization gives $\lambda_c=5.1$.
    For the unpolarized system confined in a harmonic trap with $N_{\uparrow}=N_{\downarrow}=1$, the SDFT based on BALSDA gives the threshold
    interaction strength at $\lambda_c=5.1$, while the exact diagonalization gives $\lambda_c=3.4$.
\end{thebibliography}
\end{document}